OPEN

# Isotopic effect on collisional widths and shifts of Hg clock transition induced by cold Rb atoms

Renu Bala[1,3]✉, Adam Linek[1,3], Marcin Witkowski[1], Piotr S. Żuchowski[1], Michał Zawada[1], Paul S. Julienne[2] & Roman Ciuryło[1]

**We study the isotopic dependence of collisional widths and shifts of the Hg clock transition $^1S_0$–$^3P_0$ perturbed by the Rb atoms in the temperature range from 1 nK to 1 K. For this purpose, we model the Born-Oppenheimer effective interaction potential by including the leading long-range van der Waals coefficients. For elastic collisions, we show the connection between line shape parameters in the μK temperature range and scattering lengths in ground and excited states as a function of reduced mass of the colliding Hg and Rb atoms. We confront the full quantum scattering calculations with a semi-classical approximation for collisional widths and shifts. We show that the shape resonances in excited and ground scattering states lead to significant variations of collisional line shape parameters with the change of the reduced mass of colliding atoms. We also indicate the possible influence of inelastic collisions, which could lead to universal behavior and significantly affect the dependence of collisional broadening and shifting on the isotopic combination of colliding atoms.**

In this paper, we focus on model-based studies of collisional widths and shifts of an optical clock transition affected by elastic collisions. We show a strong dependence of the transition width and shift on the reduced mass of the colliding partners, when collisions are elastic. The study uses as an example the $^1S_0$–$^3P_0$ Hg transition perturbed by Rb atoms.

The performance of optical atomic clocks exhibits a continuous improvement, surpassing that of atomic fountain clocks reliant on microwave transitions[1–14]. Numerous optical transitions have now gained recognition as secondary representations of the International System of Units (SI) for measuring time[15], thus paving the way for developing a new definition of the SI second based on optical transitions in the near future[16]. Optical atomic clocks demonstrate remarkable precision and serve as powerful tools for investigating the fundamental laws of physics. By precisely probing the stability of physical constants and, consequently, fundamental interactions such as electro-weak and -strong interactions, these clocks offer valuable insights independent of cosmological models and associated assumptions[1,17–23].

Among species used in the optical atomic clocks, the alkaline-earth and alkaline-earth-like atoms are of special importance due to the presence of the forbidden $^1S_0$–$^3P_0$ transition. The absence of a magnetic moment in the $^1S_0$ and the $^3P_0$ states makes this transition suitable for various scientific applications, including quantum simulation[24–28], computation[29–32], and gravitational wave detection[33–35]. Having the lowest susceptibility[36] to blackbody radiation shifts in group II, Hg is a candidate for the best optical frequency standard[37,38]. Additionally, the relatively strong intercombination transition $^1S_0-{}^3P_1$[39–41] allows for laser cooling of Hg with a Doppler limit at 31 μK[42–46]. Hg is also attractive due to its rich isotopic diversity, which, in combination with the forbidden transitions, $^1S_0-{}^3P_0$ and $^1S_0-{}^3P_2$, provides an opportunity for probing new fundamental interactions like Higgs boson couplings to the electron and the up and down quarks via the King plot linearity measurement[40,47–49], which may pave the way for experimental verification of the Standard Model.

The evaluation of systematic effects in single-species optical clock systems has become well-established[15,50]. Even better accuracy can be reached by utilizing two-species common-trap atomic clocks[51]. In a two-species atomic clock, one of the species is used as a reference oscillator, and the other is used as a sensor to measure perturbing factors, such as the magnetic field and the blackbody radiation. A good candidate for this system is the Hg-Rb mixture[42], with the $^1S_0-{}^3P_0$ Hg transition being a frequency standard, and two-photon 5S−7S[52] or 5S−5D[53] Rb transitions a diagnostic tool.

This paper highlights the new position in such a clock accuracy budget, which is the interspecies cold-cold atomic collision shift. The collisions between the Hg and Rb atoms lead to the broadening and shifting of the $^1$

[1]Institute of Physics, Faculty of Physics, Astronomy and Informatics, Nicolaus Copernicus University, Grudziądzka 5, 87-100 Toruń, Poland. [2]Joint Quantum Institute, University of Maryland and NIST, College Park, MD 20742, USA. [3]Renu Bala and Adam Linek have contributed equally to this work. ✉email: renub@umk.pl

$S_0-{}^3P_0$ Hg clock transition. The broadening may increase the statistical uncertainty, while the shift introduces the unknown systematic error. Here, we theoretically analyze the collisional effects in the Hg-Rb mixture. We have found the non-trivial behaviors of the isotopic effects at temperature ranges from μK up to mK, when only elastic collisions between Hg and Rb are assumed. We show that there are some isotopic pairs for which collisional effects are minimal, making these pairs suitable for two-species atomic clocks. While our work is focused on elastic collisions, it should be noted that isotopic variations of collisional broadening and shifting can change when inelastic collisions come into play. In case of Hg clock transition perturbed by Rb atoms, the Penning ionization[54] can contribute to the inelasticity of the collisions. A detailed study of this effect is beyond the scope of our work.

## Collisional width and shift
### Quantum approach
Interactions between a clock Hg atom in either ground $g$ (${}^1S_0$) or excited $e$ (${}^3P_0$) states, and a perturbing Rb atom in its ground electronic state (${}^2S_{1/2}$) can be described with good approximation by spherically symmetric single potentials $V_g(r)$ and $V_e(r)$, where $r$ is the separation between interacting atoms. These collisions can be described with a single-channel approach using one-dimensional scattering matrices $S_g(l,E)$ and $S_e(l,E)$, where $l=0,1,2,\ldots$ represents the $s,p,d,\ldots$ partial waves, respectively, and $E$ is the collision energy[55,56].

The quantum scattering cross-section $\sigma_{\text{qu}}(E)$ for collisional width and shift of the clock transition is then expressed by an infinite series over partial waves[57,58]

$$\sigma_{\text{qu}}(E)=\frac{\pi}{k^2}\sum_{l=0}^{\infty}(2l+1)\left(1-S_e(l,E)S_g^*(l,E)\right), \tag{1}$$

where $k=\sqrt{2\mu E/\hbar^2}$ is the wave number, $\hbar$ is the Planck constant $h$ divided by $2\pi$, and $\mu=m_1m_2/(m_1+m_2)$ is the reduced mass of two colliding atoms with masses $m_1$ and $m_2$. The scattering matrices $S_i(l,E)$, where $i=e$ or $i=g$, are calculated by solving radial Schrödinger equation with effective potentials $U_i(r,l)=V_i(r)+B(r)l(l+1)$ where $B(r)=\hbar^2/(2\mu r^2)$. The centrifugal part $B(r)l(l+1)$ of the effective potential plays a dominant role at low-collision energies $E$. At ultra-low energies, the contribution of higher partial waves to $\sigma_{\text{qu}}(E)$ is negligible, and only the $s$ partial wave determines the scattering cross-section.

Collisional width $\Gamma(E)$ and shift $\Delta(E)$ of the clock transition in the impact approximation is proportional to the number density $n$ of perturbing atoms and are given in circular frequency $\omega$ by[57]

$$\Gamma(E)+i\Delta(E)=n\frac{\hbar k}{\mu}\sigma_{\text{qu}}(E). \tag{2}$$

The results in this paper are presented in ordinary frequencies related to circular frequencies by $\nu=\omega/2\pi$.

In the general case for ultra-cold $s$-wave collisions complex scattering lengths can be defined as

$$\tilde{a}_i=a_i-ib_i=\lim_{k\to 0}\frac{1-S_i(l=0,E)}{2ik}. \tag{3}$$

The imaginary part of this quantity $-\text{Im}\,\tilde{a}_i=b_i$ describes inelasticity of the collisions. For pure elastic collisions, $b_i=0$, the scattering lengths $a_i$ are the major parameters that describe the ultra-cold $s$-wave collision[59]. These quantities can be used to approximate the scattering matrices for the $s$-wave at ultra-low-energy collisions[60,61]

$$S_i(l=0,E)\approx e^{-2ik\tilde{a}_i}=e^{-2ik(a_i-ib_i)} \tag{4}$$

and the term

$$1-S_e(l=0,E)S_g^*(l=0,E)\approx 1-e^{-2ik(\tilde{a}_e-\tilde{a}_g^*)}=1-e^{-2ik(a_e-a_g)}e^{-2k(b_e+b_g)}. \tag{5}$$

Therefore, Eq. (2) can be approximated as[62]

$$\Gamma(E)+i\Delta(E)\approx n\frac{\pi\hbar}{\mu k}\left(1-e^{-2ik(a_e-a_g)}e^{-2k(b_e+b_g)}\right). \tag{6}$$

Equation (6) can be further approximated by applying series expansion of exponential function up to the second order in terms of $k$

$$\Gamma(E)+i\Delta(E)\approx n\frac{\pi\hbar}{\mu k}\left(2k(b_e+b_g)-2k^2(b_e+b_g)^2+2k^2(a_e-a_g)^2+2ik(a_e-a_g)-4ik^2(b_e+b_g)(a_e-a_g)\right). \tag{7}$$

The terms of Eq. (7), linear in $k$ in the bracket, are analogues to those used in Refs.[62,63]. In the case of pure elastic collision, $b_e=b_g=0$, Eq. (7) takes the following simplified form:

$$\Gamma(E)+i\Delta(E)\approx n\frac{\pi\hbar}{\mu k}\left(2k^2(a_e-a_g)^2+2ik(a_e-a_g)\right). \tag{8}$$

## Classical approximation and thermal average

At higher temperatures, a classical approximation with straight-line trajectories can be used[64,65]. In this approximation, the time-dependent separation between the colliding atoms is $r(t) = \sqrt{\rho^2 + v^2t^2}$, where $\rho$ and $v = \hbar k/\mu$ represent the impact parameter and the relative velocity of the colliding atoms, respectively. The sum over quantum partial waves, given in Eq. (1), is replaced by integration over the classical impact parameter $\rho$. The quantum scattering cross-section $\sigma_{\rm qu}$ is approximated by the classical cross-section

$$\sigma_{\rm cl}(E) = 2\pi \int_0^\infty d\rho \; \rho \left(1 - S_e(\rho, E) S_g^*(\rho, E)\right), \tag{9}$$

where

$$S_i(\rho, E) = e^{-i\phi_i(\rho, v)} \tag{10}$$

and

$$\phi_i(\rho, v) = \frac{1}{\hbar} \int_{-\infty}^{+\infty} dt V_i(r(t)). \tag{11}$$

Finally, collisional width and shift in the classical limit can be written as

$$\Gamma(E) + i\Delta(E) = nv\sigma_{\rm cl}(E). \tag{12}$$

The thermally averaged collisional width and shift in non-degenerate thermal gas at temperature $T$ can be calculated by averaging Eqs. (2) and (12) over the Maxwell-Boltzmann distribution[65,66]

$$\Gamma + i\Delta = \frac{2}{\sqrt{\pi}} \int_0^\infty \frac{dE\sqrt{E}}{(k_BT)^{3/2}} \, e^{-E/(k_BT)} \left(\Gamma(E) + i\Delta(E)\right), \tag{13}$$

where $k_B$ is Boltzmann constant.

## Scattering length for van der Waals interaction

Assuming van der Waals interactions, in the zero-energy limit for collisions, the *s*-wave scattering length with a quantum correction to the Wentzel-Kramers Brillouin (WKB) approximation can be given by[67]

$$a = \bar{a}\left[1 - \tan\left(\Phi - \frac{\pi}{8}\right)\right], \tag{14}$$

where $\bar{a}$ is the average scattering length and $\Phi$ is the semiclassical phase. The average scattering length is equal to

$$\bar{a} = 2^{-3/2} \frac{\boldsymbol{\Gamma}(3/4)}{\boldsymbol{\Gamma}(5/4)} \left(\frac{2\mu C_6}{\hbar^2}\right)^{1/4}, \tag{15}$$

where $C_6$ represents the long-range interaction van der Waals coefficient, and here $\boldsymbol{\Gamma}(\ldots)$ represents the gamma Euler function. The semiclassical phase $\Phi$ can be expressed by the interaction potential $V(r)$ between colliding atoms as

$$\Phi = \frac{\sqrt{2\mu}}{\hbar} \int_{r_0}^\infty dr \sqrt{-V(r)}, \tag{16}$$

where $r_0$ is the inner classical turning point. The semiclassical phase $\Phi$ determines the number of bound vibrational states $N$ with zero orbital angular momentum[68]

$$N = \left[\frac{\Phi}{\pi} + \frac{3}{8}\right]. \tag{17}$$

## Modeling Penning ionization for *s*-wave collisions

During the collisions of excited Hg atoms with Rb atoms in their ground state, Penning ionization is possible. We estimate the possible influence of Penning ionization on collisional broadening and shifting of the Hg clock transition by adapting the model developed by Idziaszek and Julienne[69] for reactive collisions of ultra cold molecules. To apply this model for *s*-wave collisions of excited Hg atoms with Rb atoms in the ground state, we use the scattering matrices $S_e^{\rm el}(l = 0, E)$ for the *s*-wave, which describes a pure elastic collision and convert it into "energy dependent" scattering length as

$$a_e(E) = \frac{1}{ik} \frac{1 - S_e^{\rm el}(l = 0, E)}{1 + S_e^{\rm el}(l = 0, E)}. \tag{18}$$

Next, we insert such determined $a_e(E)$ into equation derived in Ref.[69]

$$\tilde{a}_e(E) = a_e(E) + \bar{a}_e y \frac{1+(1-s)^2}{i+y(1-s)^2}\ , \tag{19}$$

where $s = a_e(E)/\bar{a}_e$ and $y$ is a dimensionless parameter describing the reactive or inelastic part of the short-range collision. The parameter $0 \le y \le 1$ is related to the probability of irreversible loss of the incoming scattering flux from the entrance channel due to the dynamics at short range. For $y=1$ there is no outgoing flux and $\tilde{a}_e(E) = \bar{a}_e - i\bar{a}_e$ gets a universal form. Whereas for $y$ between unity and zero there is some back reflection. Finally, for $y=0$ (no losses), the incident and reflected fluxes are equal, and $\tilde{a}_e(E) = a_e(E)$ is simply the elastic scattering length. We can now use $\tilde{a}_e(E)$ to model scattering matrix which includes the inelasticity of collisions in the short range part of collisions

$$S_e^{\mathrm{inel}}(l=0,E) = \frac{1-ik\tilde{a}_e(E)}{1+ik\tilde{a}_e(E)}\ . \tag{20}$$

This modified scattering matrix for the excited state, Eq. (20), is used in Eq. (1) to estimate the possible impact of Penning ionization on collisional broadening and shifting of clock transition in the μK range of gas temperature.

Writing the scattering matrices for ground and excited states in the form analogous to Eq. (20), results in an approximate expression for collisional width and shift given below

$$\Gamma(E) + i\Delta(E) \approx n\frac{\pi\hbar}{\mu k}\frac{(2k^2(\tilde{a}_e(E)-a_g(E))^2 + 2ik(\tilde{a}_e(E)-a_g(E))(1+k^2\tilde{a}_e(E)a_g(E))}{(1+k^2\tilde{a}_e^2(E))(1+k^2a_g^2(E))}\ . \tag{21}$$

For the derivation of the above expression, we have assumed that $\tilde{a}_e(E)$ can be a complex, but $a_g(E)$ is a real one. In a more general case, where inelasticity of ground state collisions is also taken into account, ground state scattering length should be complex and $a_g(E)$ should be replaced by $\tilde{a}_g^*(E)$ in Eq. (21). In case where both excited and ground state collisions are elastic ($y=0$, $\tilde{a}_e(E)$ and $a_g(E)$ are real) by limiting the Taylor expansion of the second fraction of Eq. (21) to terms proportional to $k$ and $k^2$, the above expression can be reduced to Eq. (8). We also checked that in case of $s$-wave elastic collisions at energy $E/k_B = 1$ μK, Eq. (21) with inserted scattering lengths from Fig. 1(a) and 1(b) agrees very well with full quantum scattering calculations (agreement better than 1% for the maximum value). Both results would be indistinguishable on Figs. 2 and 3. For the collision energy $E/k_B = 10$ μK, agreement is slightly worse, but still better than 7%.

## Interaction potential

The Born-Oppenheimer interaction potential calculated using state-of-the-art electronic structure methods for the ground state of Hg–Rb is known with an accuracy of about 20 $\mathrm{cm}^{-1}$ (about 5%). Currently, the shape of the interaction potential for the excited electronic state with the asymptote $^3\mathrm{P}_0 - ^2\mathrm{S}_{1/2}$ is unknown. Calculating such a potential is extremely challenging since its asymptote lies in the continuum above the ionization threshold of Rb. Thus, we introduce the model potentials $V_i(r)$ in the Lennard-Jones form as follows[70,71]:

$$V_i(r) = -\frac{C_{6,i}}{r^6}\left(1 - \frac{\sigma_i^6}{r^6}\right) \tag{22}$$

with the leading long-range van der Waals coefficient $C_{6,i}$. The short-range parameter $\sigma_i$ describes the distance at which the potential $V_i(r)$ is zero. Typically, the *ab initio* calculated potentials, with only a few exceptions[72–75], do not allow determination of the scattering length. Therefore, we set the value of the scattering length for one arbitrarily chosen pair of colliding isotopes, $^{202}$Hg + $^{87}$Rb.

To describe the interaction potential in the electronic ground state of the HgRb system, we use the value of $C_{6,g} = 949.7\ E_h a_0^6$, where $a_0$ and $E_h$ represent the Bohr radius and the Hartree energy[76,77], respectively, and the number of bound vibrational states $N_g = 44$[78]. We assume that the value of scattering length for the ground state of $^{202}$Hg + $^{87}$Rb is equal to the average scattering length Eq. (15) for van der Waals interaction $\bar{a}_g = 57.57\ a_0$. This assumption, together with $C_{6,g}$ and $N_g$, yields $\sigma_g = 6.5604\ a_0$ for the ground electronic state setting the proper quantum defect[79,80]. The values $C_{6,g}$ and $\sigma_g$ were used to calculate the scattering properties for all pairs of isotopes in this state. The $C_6$ coefficients are proportional to the polarizabilities of interacting species. Therefore, for the excited $|e\rangle$ state of the HgRb molecule with the dissociation asymptote $^{202}$Hg ($^3\mathrm{P}_0$) + $^{87}$Rb ($^2\mathrm{S}_{1/2}$), we estimate the value of $C_{6,e} \approx 1500\ E_h a_0^6$ by scaling the $C_{6,g}$ approximately 1.5 times using the known values of $C_{6,i}$ from the SrRb system with the asymptotes Sr ($^1\mathrm{S}_0$) + Rb ($^2\mathrm{S}_{1/2}$) and Sr ($^3\mathrm{P}_0$) + Rb ($^2\mathrm{S}_{1/2}$)[81,82]and the static atomic polarisabilities[83–85]

$$C_{6,e} \approx \frac{\alpha(\mathrm{Hg}\ ^3\mathrm{P}_0)}{\alpha(\mathrm{Hg}\ ^1\mathrm{S}_0)}\frac{\alpha(\mathrm{Sr}\ ^1\mathrm{S}_0)}{\alpha(\mathrm{Sr}\ ^3\mathrm{P}_0)}\frac{C_{6,e}(\mathrm{SrRb})}{C_{6,g}(\mathrm{SrRb})}C_{6,g}, \tag{23}$$

where $\alpha(\mathrm{Hg}\ ^1\mathrm{S}_0) = 33.91(34)\ a_0^3$ and $\alpha(\mathrm{Sr}\ ^1\mathrm{S}_0) = 197.2(2)\ a_0^3$[83,86] are the static polarisabilities in the ground state of Hg and Sr, respectively, and $\alpha(\mathrm{Hg}\ ^3\mathrm{P}_0) = 81.97\ a_0^3$[84] and $\alpha(\mathrm{Sr}\ ^3\mathrm{P}_0) = 410(28)\ a_0^3$[85] are the static polarisabilities in the excited state of Hg and Sr, respectively. The long-range van der Waals coefficient in the ground electronic state $C_{6,g}(\mathrm{SrRb})$ = 3755.7164 $E_h a_0^6$[81]. To calculate the value of $C_{6,e}(\mathrm{SrRb})$ = 4857.64 $E_h a_0^6$ for the excited electronic state with the asymptote Sr ($^3\mathrm{P}_0$) + Rb ($^2\mathrm{S}_{1/2}$) we diagonalize the matrix given by Eq. (8) in Ref.[82] using energies of the excited states $^3P_2$ and $^3\mathrm{P}_0$ in atomic Sr[87]. We have also estimated the value of $C_6$ for the excited state of the HgRb system by the Slater-Kirkwood formula using the static polarizabilities as[88]

$$C_6 = \frac{3}{2}\frac{\alpha(\mathrm{Hg}\ ^3\mathrm{P}_0)\alpha(\mathrm{Rb}\ ^2\mathrm{S}_{1/2})}{\sqrt{\frac{\alpha(\mathrm{Hg}\ ^3\mathrm{P}_0)}{n_{Hg}}}+\sqrt{\frac{\alpha(\mathrm{Rb}\ ^2\mathrm{S}_{1/2})}{n_{Rb}}}}E_h a_0^{3/2}, \tag{24}$$

where $n_{Hg} = 2$ and $n_{Rb} = 1$ are the number of electrons in the outer shell of Hg and Rb atoms, respectively, $\alpha$ (Hg $^3$P$_0$) $= 81.97a_0^3$[84] and $\alpha$(Rb $^2$S$_{1/2}$) $= 319.8(0.3)a_0^3$[83]. The value of $C_6$ estimated with the above formula is $1619.15\ E_h a_0^6$, which differs from the value used in the current work by less than 8%. The value $\sigma_e = 6.3069\ a_0$ is set so the interaction potential $V_e(r)$ supports the number of bound vibrational states, $N_e = 60$, and the value of the scattering length in the excited state for the $^{202}$Hg + $^{87}$Rb isotopic pair is equal to 5.26 $a_0$. The chosen value of the scattering length is close to one-tenth of the average $\bar{a}_g$. To address the uncertainty of the used potentials' parameters, we perform our studies for a broad range of reduced masses of the Hg–Rb system, which is equivalent to the change of the potential well, as the WKB phase shift, Eq. (16), depends on product $\mu$ and $V(r)$.

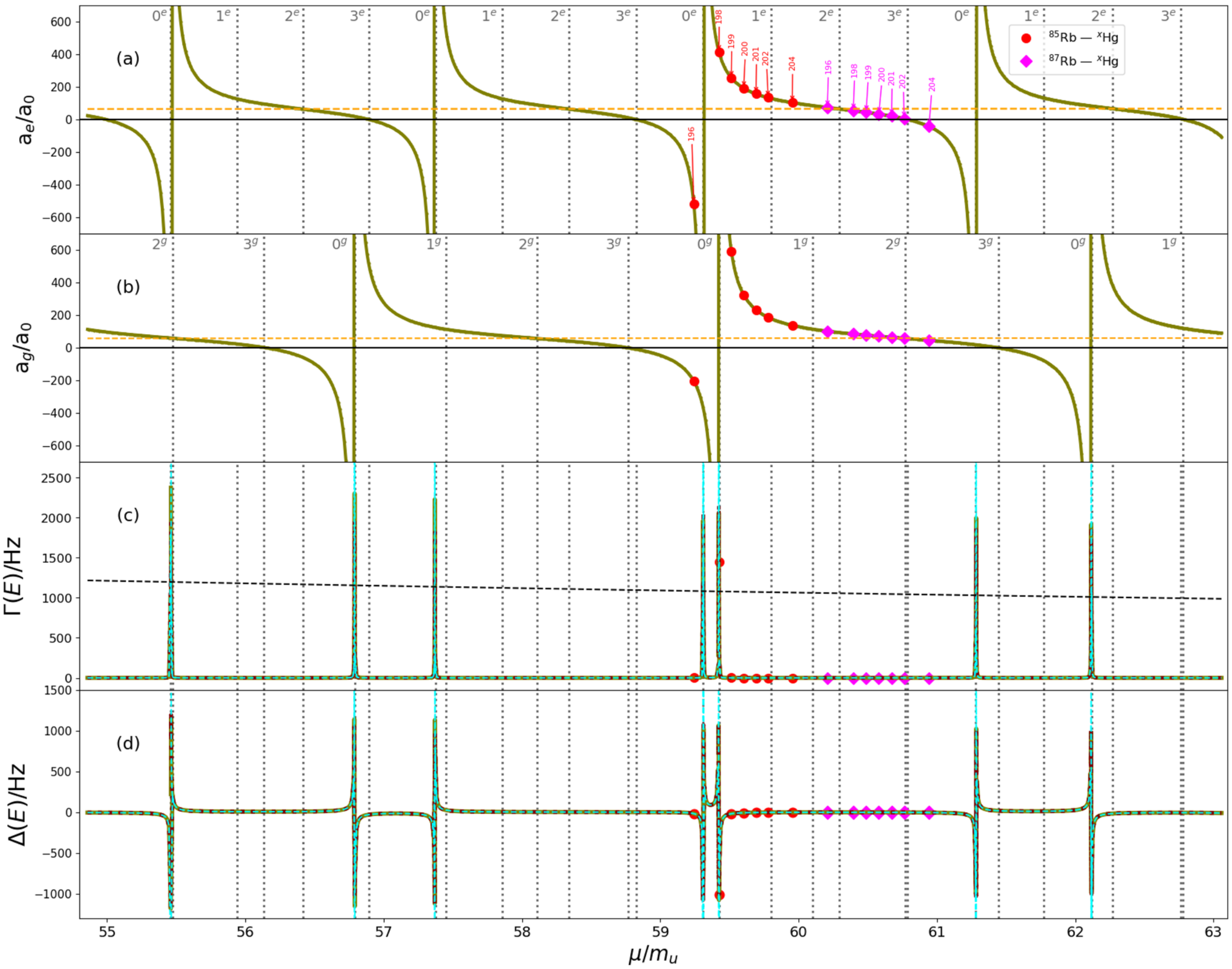

**Fig. 1**. Two upper panels show the *s*-wave scattering lengths for the collisions of Hg and Rb atoms: (**a**) Hg atom is in its excited ($^3$P$_0$) state and Rb atom is in its ground ($^2$S$_{1/2}$) state; (**b**) both Hg and Rb atoms are in their ground states, ($^1$S$_0$) and ($^2$S$_{1/2}$), respectively. The dotted orange horizontal lines in the top two plots represent the average scattering lengths $\bar{a}_e$ ranges between 62.9 $a_0$ to 65.1 $a_0$ and $\bar{a}_g$ between 56.1 $a_0$ to 58.1 $a_0$, for the excited and ground state, respectively. The value of the scattering length of $^{198}$Hg + $^{85}$Rb in the ground state is out of the scale; hence, it is not shown in the figure. Two lower panels present the collisional width (**c**) and shift (**d**) calculated for collisional energy $E/k_B = 10^{-9}$ K and Rb number density $n = 10^{12}$ cm$^{-3}$. The solid olive lines represent the full quantum scattering calculations. The results of the low-temperature approximation are shown by the dotted maroon lines and the dashed cyan lines, following Eqs. (6) and (8), respectively. The pairs of natural isotopes $^{85}$Rb-$^x$Hg and $^{87}$Rb-$^x$Hg, are represented by the red circles and magenta diamonds, respectively, where *x* = 196, 198, 199, 200, 201, 202, 204. The black dashed lines represent the results for *s*-wave scattering in the unitary limit, i.e. $S_e(0,E) = 0$. The vertical dotted grey lines represent the position of the shape resonances labeled with $l^{(e,g)}$, where $l = 0, 1, 2, 3, \ldots$ represent the *s*, *p*, *d*, *f*, ... partial waves, and the superscript (*e*, *g*) defines the electronic state.

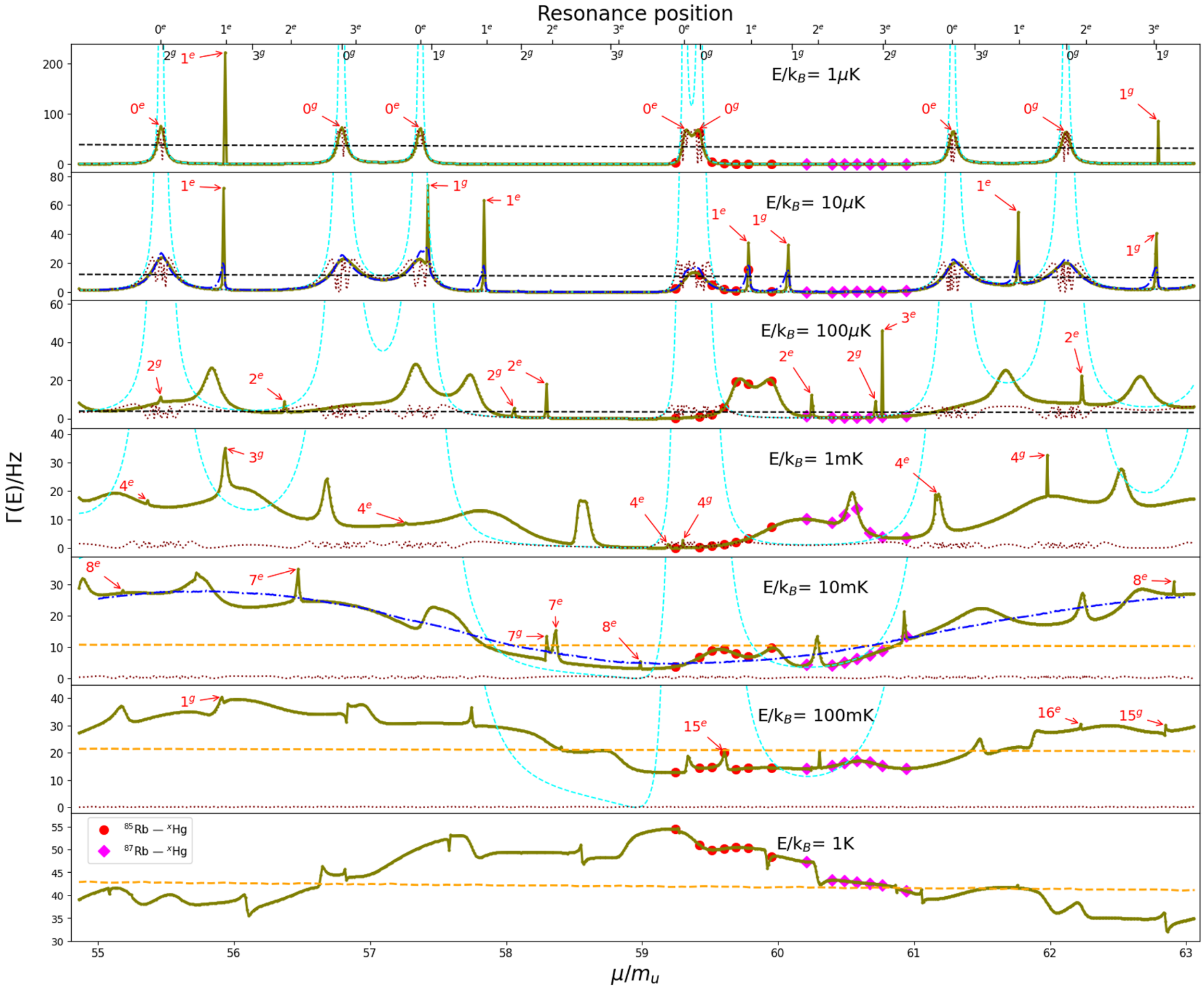

**Fig. 2**. The figure shows the variation of collisional widths against the reduced mass of the Hg-Rb system in the temperature range from 1 μK to 1 K. The solid olive lines represent the full quantum scattering calculations. The dashed-dotted blue lines represent the thermally averaged full quantum scattering calculations at temperatures 10 μK and 10 mK. The results of the low-temperature approximation are shown by the dotted maroon lines and the dashed cyan lines, using Eqs. (6) and (8), respectively. The dashed orange lines represent the classical limit calculations. The pairs of natural isotopes $^{85}$Rb-$^{x}$Hg and $^{87}$Rb-$^{x}$Hg, are represented by the red circles and magenta diamonds, respectively, where $x$ = 196, 198, 199, 200, 201, 202, 204. The black dashed lines represent the results for $s$-wave scattering in the unitary limit, i.e. $S_e(0,E)=0$ . The shape resonances are labelled with $l^{(e,g)}$, where $l=0,1,2,3,\ldots$ represent the $s$, $p$, $d$, $f$... partial waves, and the superscript ($e$, $g$) defines the electronic state. The results were computed for Rb number density $n=10^{12}$ cm$^{-3}$.

## Quantum scattering calculations

We have numerically solved the Schrödinger equation with the renormalized Numerov method to calculate the single channel bound and scattering states wavefunctions for given interaction potentials[89,90]. The bound state calculations allowed us to adjust the potential parameters to match the assumed number of bound states.

The single channel $S$-matrices were determined from asymptotic scattering wavefunctions[91]

$$F_i(l,E;r) \underset{r\to\infty}{=} A_i(l,E)\,(J(l,E;r)+K_i(l,E)N(l,E;r)) \tag{25}$$

expressed in terms of $J(l, E; r)$ and $N(l, E; r)$, which are related to spherical Bessel functions. In Eq. (25), $A_i(l,E)$ is a normalization factor and $K$-matrices can be given in the following form:

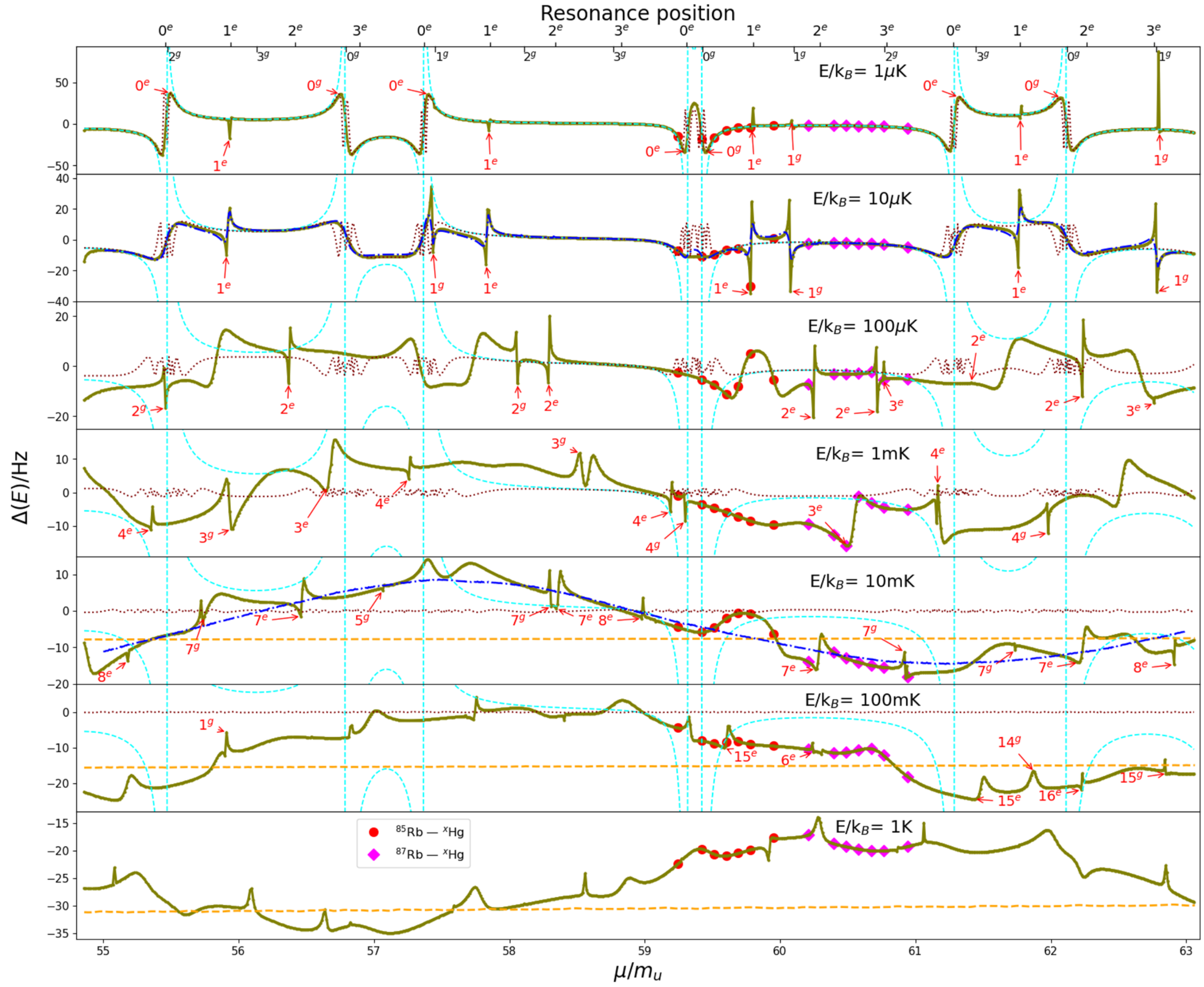

**Fig. 3**. The figure shows the variation of collisional shifts against the reduced mass of the Hg-Rb system in the temperature range from 1 μK to 1 K. The solid olive lines represent the full quantum scattering calculations. The dashed-dotted blue lines represent the thermally averaged full quantum scattering calculations at temperatures 10 μK and 10 mK. The results of the low -temperature approximation are shown by the dotted maroon lines and the dashed cyan lines, following Eq. (6) and Eq. (8), respectively. The dashed orange lines represent the classical limit calculations. The pairs of natural isotopes $^{85}$Rb-$^x$Hg and $^{87}$Rb-$^x$Hg, are represented by the red circles and magenta diamonds, respectively, where $x = 196, 198, 199, 200, 201, 202, 204$. The shape resonances are labelled with $l^{(e,g)}$, where $l = 0, 1, 2, 3, \ldots$ represent the *s*, *p*, *d*, *f*... partial waves, and the superscript (*e*, *g*) defines the electronic state. The results were computed for Rb number density $n = 10^{12}$ cm$^{-3}$.

$$K_i(l, E) = W(F_i, J)(W(N, F_i))^{-1}, \tag{26}$$

where

$$W(X, Y) = X(r)\frac{\partial Y(r)}{\partial r} - \frac{\partial X(r)}{\partial r}Y(r) \tag{27}$$

is Wronskian between functions *X*(*r*) and *Y*(*r*). Equation (26) is obtained using relation $W(N, J) = 1$. Furthermore, the open channel scattering *S*-matrices are calculated from *K*-matrices[91]

$$S_i(l, E) = (1 + iK_i(l, E))\,(1 - iK_i(l, E))^{-1}, \tag{28}$$

and can be expressed with the Wronskian between the numerically determined wavefunctions $F_i(l, E; r)$ and functions *J*(*l*, *E*; *r*) and *N*(*l*, *E*; *r*) at large enough separations *r*

$$S_i(l, E) = \left(W(N, F_i)^2 - W(F_i, J)^2 + 2iW(N, F_i)W(F_i, J)\right)\left(W(N, F_i)^2 + W(F_i, J)^2\right)^{-1}. \tag{29}$$

## Results and discussion

In two upper panels of Fig. 1, we present the variation of the elastic scattering lengths for the ground and excited states as a function of the reduced Hg and Rb system mass. We estimate the scattering lengths in the zero collisional energy limit approximated by the scattering energy of $E/k_B = 10^{-15}$ K. The values of the calculated scattering lengths reveal singularities (*s*-wave scattering resonances), which, according to Eq. (14), are almost periodical. The span of the reduced mass of existing Rb and Hg isotopes is comparable to the distance between the adjacent resonances of the scattering lengths in the excited state. This is not the case for the ground state, where the distance between the resonances is much larger. This results from the fact that the spacing between adjacent scattering length resonances, $\Delta\mu$, can be approximated by the simple expression $\Delta\mu \approx 2\mu/N$[71] and the ratio $N_e/N_g$, representing the assumed number of bound states in the excited and ground electronic states, is approximately 1.36. Furthermore, it is worth noting that Bohn and Wang[92] have recently discussed the probability distribution of scattering lengths.

For better understanding of the results presented in this section, Fig. 1 includes the vertical dotted grey lines, indicating the positions of partial waves *p*, *d*, and *f* shape resonances determined for ultra-cold collisions. According to Gao[93,94], who used the framework of quantum defect theory for a van der Waals system, *g*-wave resonance should coincide with *s*-wave resonance, and *p*, *d*, *f*-wave resonances should be equally spaced between *s* and *g*-wave resonances. Notably, *d*-wave resonance coincides with scattering length equal to $\bar{a}$. See also a discussion related to ultra-cold collisions of Yb atoms by Borkowski *et.al.*[95].

In general, the absolute positions of the *s*-wave resonances in the ground and excited state and the distance between them have not been measured and are unknown yet. The absolute position of the *s*-wave resonances in the graphs presented in Fig. 1 depends on the given quantum defect and the distance between *s*-wave resonances depends on the assumed number of bound states. Therefore, the calculated graphs can be moved left or right in Fig. 1. In our work, the position and span of the resonances are controlled by the $\sigma$ parameter in the potential. Nevertheless, a comparison of the plotted graphs with the known distribution of reduced mass in Hg-Rb pairs shows that under the real experimental conditions, there will be colliding isotopic pairs with both small and very large values of the scattering length.

The two bottom panels of Fig. 1 show the dependence of the width and shift of clock transition induced by elastic collisions on the reduced isotopic mass at a collision energy of $E/k_B = 10^{-9}$ K. The resonance features of collisional width and shift correspond to the scattering length resonances in the ground or excited states. For ultra-cold collisions, only the *s* partial wave participates in the scattering cross-section due to the centrifugal barrier preventing higher partial waves from contributing. For this energy scale, we verified the applicability of the Eqs. (6) and (8) by comparing their predictions with the full quantum scattering calculations, getting an agreement better than 0.2% and 0.4%, respectively, apart from the resonances.

It can be seen that, for example, for the assumed specific values of $C_{6,i}$ and $\sigma_i$ used in the presented calculations, the collisional width and shift corresponding to the reduced mass of $^{198}$Hg and $^{85}$Rb system are substantially large, $1.45 \times 10^3$ Hz and $-1.01 \times 10^3$ Hz, respectively, with respect to other isotopes. This is because of the large value of the scattering length in the ground state ($\approx 55000\ a_0$). In contrast, for the isotopic pair $^{198}$Hg$^{87}$Rb having similar scattering lengths in the ground and excited states, 83.8 $a_0$ and 53.9 $a_0$, respectively, the collisional width and shift are a few orders of magnitude smaller, $1.31 \times 10^{-3}$ Hz and -1.66 Hz, respectively. For different values of $C_{6,i}$ and $\sigma_i$, other isotope pairs can hit the resonance.

The dependence of elastic collisional widths and shifts on the reduced mass are studied at different collisional energies $E/k_B$ ranging from 1 μK to 1 K as shown in Figs. 2 and 3, respectively. The reduced mass step size, which determines the resolution of the computation, varies from $0.007\ m_u$ to $0.013\ m_u$. For the lower energies, $E/k_B \leq 1$ μK, the higher partial wave resonances can be spotted only when the probed reduced mass coincides with the resonance's positions. For example, the *p*-wave resonances for collisional energy, $E/k_B = 1$ μK and for reduced masses around $55.9\ m_u$ and $62.8\ m_u$ appeared in Figs. 2 and 3. For higher collisional energies, $E/k_B > 1$ μK, the widths of resonances associated with higher partial waves become wider and clearly visible within the resolution of our computations. Additionally, the lower partial waves' structures become indistinct. At low collision energies, positions of partial wave resonances are very well predicted by the Gao theory[93,94] mentioned at the beginning of this section. However, when the collision energy increases, discrepancies start to appear. For example, the positions of *g*-wave resonances at 1 mK differ from the positions of *s*-wave resonances at 1 nK or 1 μK.

The effect of thermal averaging on collisional widths and shifts is studied for temperatures 10 μK and 10 mK and is depicted by dashed-dotted blue lines in Figs. 2 and 3. At 10 μK, the results calculated with thermal averaging do not show qualitative differences with respect to those calculated for a single collisional energy $E/k_B = 10$ μK, and the resonances remain clearly visible in both cases. In contrast, the resonances observed for collisional energy of $E/k_B = 10$ mK are indistinguishable in the results calculated with thermal averaging. The specific values of collisional widths and shifts for accessible pairs of isotopes highly depend on the positions of the *s*-wave resonances in both the ground and excited states. In the case of Hg clock transition perturbed by Rb with a number density of $10^{12}$ cm$^{-3}$ at a temperature of 10 μK, the collisional widths and shifts in Figs. 2 and 3 vary from 0 Hz to 32 Hz and −17 Hz to 21 Hz, respectively. At temperatures around 1 μK, the variation in the collisional widths and shifts is expected to be up to twice as large.

We demonstrate that the low-energy approximation given by Eqs. (6) and (8) (dotted marron and dashed cyan in Figs. 2 and 3) well reproduces the *s*-wave resonances at their wings for collisional energies, $E/k_B = 1$ μK, and lower. To validate that, we looked at the *s*-wave resonance at $\mu = 55.46\ m_u$ for collisional energy $E/k_B = 1$ μK. For the reduced mass detunings from the center of the resonance exceeding the full width at half maximum (FWHM), the low-energy approximations given by Eq. (6) and (8) exhibit agreement for the width (shift) with the full quantum scattering calculations to within 13% (5%) and 19% (18%), respectively. For collisional energies, $E/k_B \geq 10$ μK, the approximations no longer apply.

We also tested the classical approximation (dashed orange lines in Figs. 2 and 3) for collisional energies larger than $E/k_B = 10$ mK. The isotopic variation in collisional widths and shifts across the span of natural isotopes' pairs Rb-Hg ($\mu$ from 59.24 $m_u$ to 60.94 $m_u$) is below 1%, which is much smaller compared to the results of the full quantum scattering calculations. For the collisional energy, $E/k_B = 1$ K, the agreement between the full quantum scattering calculations and the classical approximation for the collisional widths and shifts is approximately 30% and 60%, respectively. The collisional widths (shifts) calculated using the full quantum scattering approach and classical approximation both averaged over the entire reduced mass range presented in Figs. 2 and 3, agree within 3% (22%).

We have also studied the variation of widths and shifts as a function of collisional energy $E/k_B$ ranging from 1 nK to 1K, as shown in Fig. 4. To show possible behaviours, we choose two pairs of isotopes: one corresponding to a very large scattering length in the ground state ($^{198}$Hg$^{85}$Rb) and the other corresponding to the similar moderate scattering lengths in ground and excited states ($^{198}$Hg$^{87}$Rb). It can be seen from Fig. 4 that the collisional broadening and shifting at low energy of collisions is extremely large for the first case $^{198}$Hg$^{85}$Rb in contrast to the second case $^{198}$Hg$^{87}$Rb, where collisional impact on the spectral line shape is small. On the other hand, as collision energy $E/k_B$ approaches 1 K, both isotope combinations have similar collisional widths and shifts, and resonance effects disappear. We need to emphasize that for the real case scenario, where $C_{6,i}$ and $\sigma_i$ can be different from those assumed in this work, the behavior discussed above could occur for other isotopic pairs.

Looking into collisional broadening and shifting of Hg clock transition perturbed by Rb atoms in the ground electronic state, two effects should be mentioned. The first is the photoionization of Rb atoms by the $^1$S$_0$–$^3$P$_0$ transition probing laser, and the second is Penning ionization of Rb atoms during collisions with Hg atoms in the excited clock state $^3$P$_0$. In the case of photoionization of Rb atoms in the ground electronic state, it was shown experimentally[96,97] that in the frequency range near the Hg clock transition, the Cooper minimum is observed, which is also supported by the theoretical prediction. This makes photoionization small enough to make possible simultaneous magneto-optical trapping of Hg and Rb atoms[42]. In the first approximation, we can ignore the influence of the $^1$S$_0$–$^3$P$_0$ transition probing laser on the Rb atoms.

Penning ionization[54] can lead to the irreversible loss of the incoming scattering flux of Hg atoms in the clock state $^3$P$_0$. To address this issue, we used the model developed by Idziaszek and Julienne[69]. For this purpose, we calculated the collisional width and shift for the collisional energy $E/k_B = 1$ μK using Eqs. (1), (2), (19) and (20), which is equivalent to use of Eq. (21). The computed results are presented in Fig. 5 and are limited to *s*-wave collisions only. We studied the dependence of collisional width and shift on the probability of irreversible loss of the incoming scattering flux from the entrance channel due to the dynamics at short range leading to Penning ionization described by a dimensionless parameter "*y*". For $y = 1$ there is no outgoing flux and collisions in excited state have an universal character described by $\tilde{a}_e(E) = \bar{a}_e - i\bar{a}_e$. In this case, "*e*"-resonances in $\Gamma(E)$ and $\Delta(E)$ associated with resonances of elastic scattering in the excited state, shown in Fig. 1(a), are no longer observed. Obviously, "*e*"-resonances appear again when $y \neq 1$. In case of $y = 0.1$ the magnitude of "*e*"-resonances are more than three times smaller compared to the pure elastic case, $y = 0$. For comparison, in

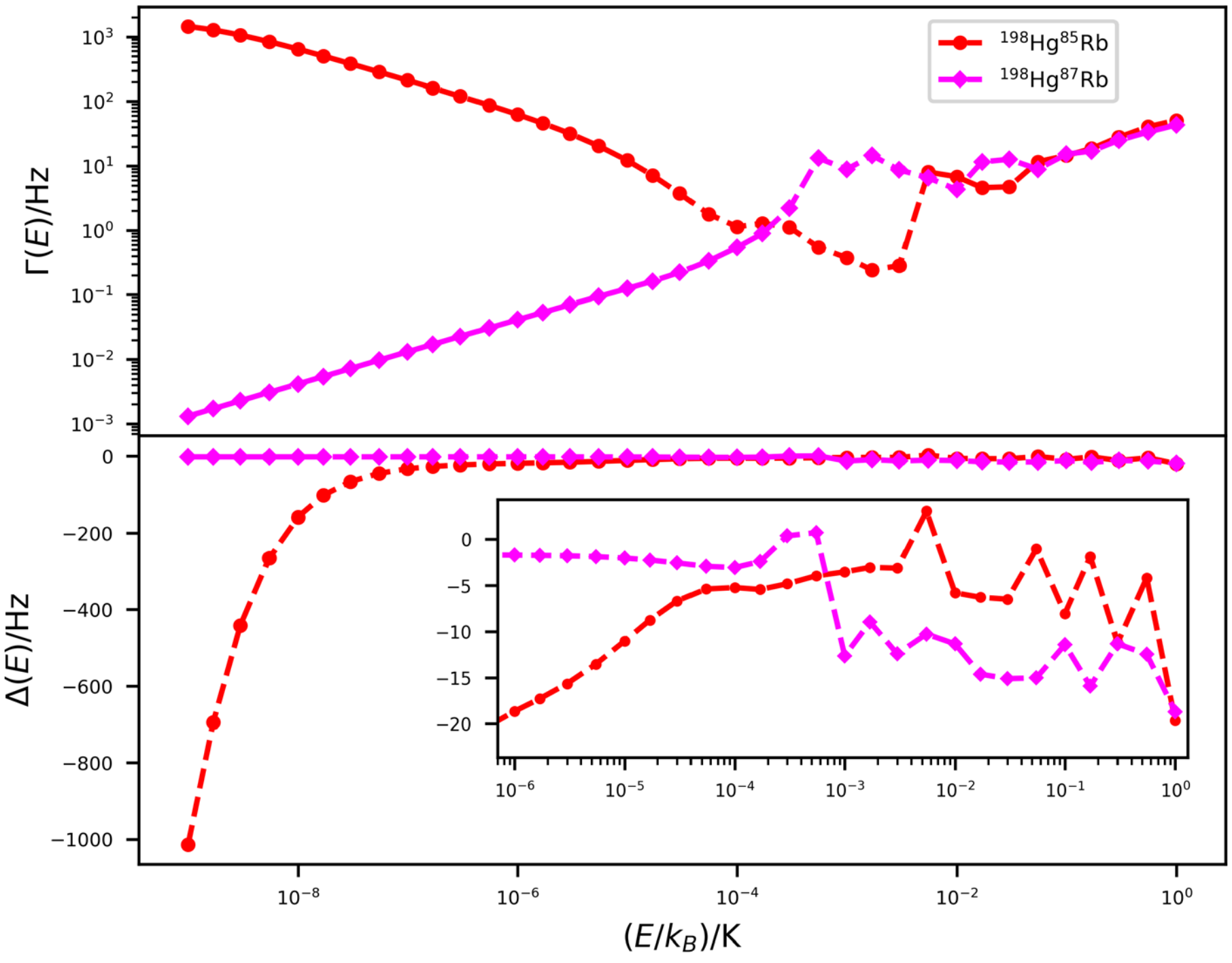

**Fig. 4**. The figure shows the variation of collisional widths and shifts against the energy of collision $E/k_B$ range from 1 nK to 1 K. The results were computed for Rb number density $n = 10^{12}$ cm$^{-3}$.

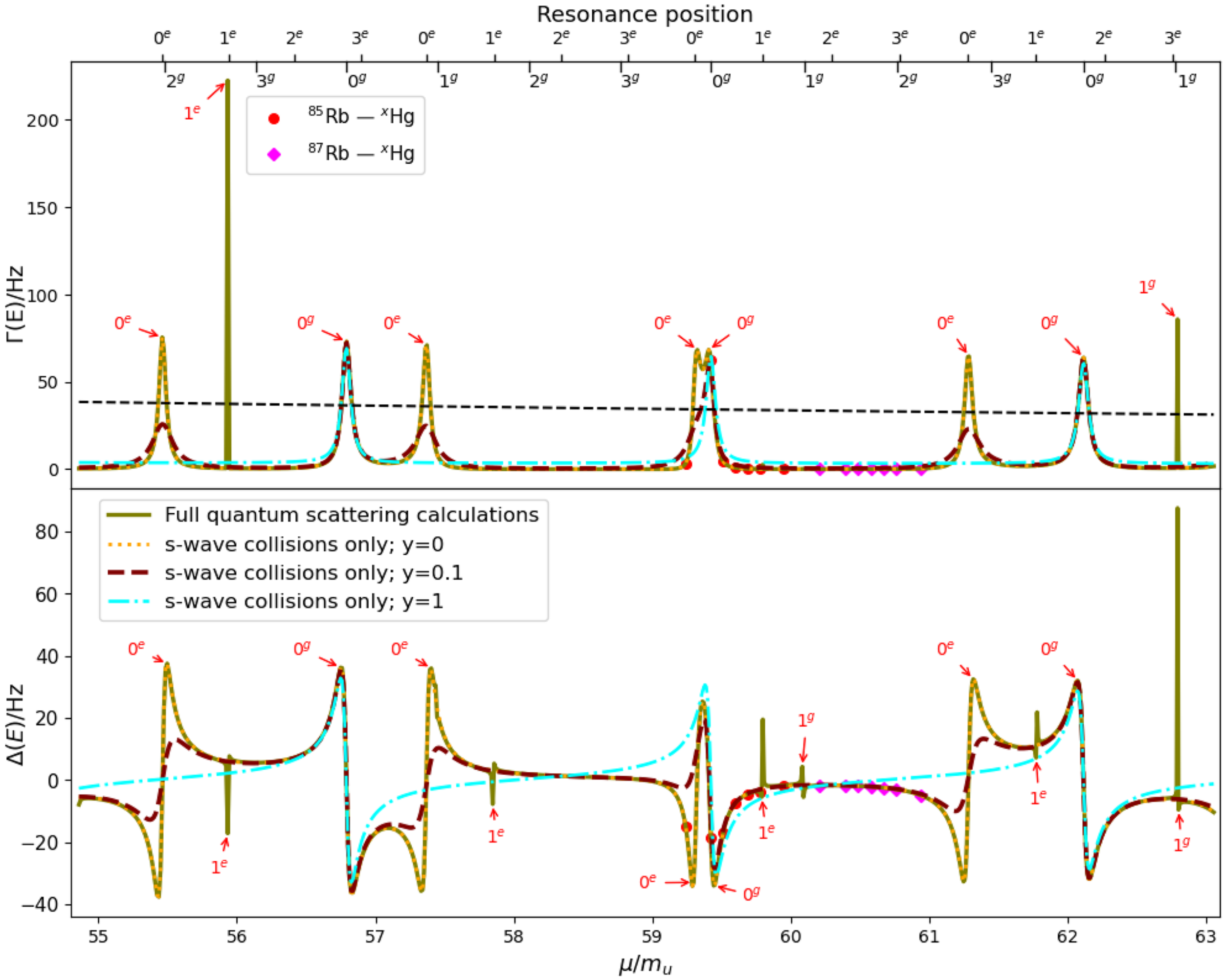

**Fig. 5**. The figure shows the variation of collisional shifts and widths against the reduced mass of the Hg-Rb system at the temperature 1 μK. The solid olive lines represent the full quantum scattering calculations. The dotted orange, dashed maroon, and dashed-dotted cyan lines correspond to the collisional parameters for s-wave collisions only for $y = 0, 0.1$, and 1, respectively. These data have been computed using Eqs. (19) and (20). The pairs of natural isotopes $^{85}$Rb-$^{x}$Hg and $^{87}$Rb-$^{x}$Hg, are represented by the red circles and magenta diamonds, respectively, where $x = 196, 198, 199, 200, 201, 202, 204$. The black dashed lines represent the results for s-wave scattering in the unitary limit, i.e. $S_e(0, E) = 0$. The shape resonances are labelled with $l^{(e,g)}$, where $l = 0, 1, 2, 3, \ldots$ represent the *s*, *p*, *d*, *f*... partial waves, respectively, and the superscript (*e*, *g*) defines the electronic state. The results were computed for Rb number density $n = 10^{12}$ cm$^{-3}$.

Figures 1 and 5, we also show results for scattering in the excited state in the unitary limit $S_e(l, E) = 0$, when Eqs. (1) and (2) are reduced to the form: $\Gamma(E) + i\Delta(E) = n\pi\hbar/(\mu k)$.

Penning ionization was studied in the regime of ultra-cold collisions[69,98,99], however, at this point, we were not able to estimate its importance for the case of Hg-Rb collisions. Therefore, in this work, we compared results obtained only for the elastic collisions with those from Idziaszek and Julienne model[69] by varying probability of reduction of the outgoing scattering flux caused by Penning ionization which takes place at short range. Clearly, the pattern of isotopic dependence in collisional broadening and shifting of Hg $^1S_0$–$^3P_0$ clock transition may serve as an experimental indicator of the importance of Penning ionization in the Hg-Rb system in the future. However, more detailed study of this issue is beyond the scope of this work.

We have shown the possibility of significant variation of collisional shift from one isotopic pair of HgRb to another, see Figs. 3 and 5, in the experimentally relevant range of temperatures from 1μK to 100μK. It could be exploited to find the isotopic pairs with the smallest possible collisional shift. For example, for elastic collisions at temperature 1μK, the smallest collisional shift (approximately −1.7 Hz) occurs for $^{196}$Hg$^{87}$Rb and $^{198}$Hg$^{87}$Rb isotopic pairs. On the other hand, when inelastic collisions are included, the collisional shift dependence on the reduced mass crosses zero within the range where isotopic pairs of HgRb exist. In this case, for $^{201}$Hg$^{87}$Rb, the calculated collisional shift is as small as +0.04 Hz. Again, we need to note that in the real case, where real interaction is different from that assumed in this paper, this behavior could occur for other isotopic pairs.

Nevertheless, Figs. 3 and 5 give an impression what kind of behavior can be expected taking into account that the particular positions of the resonances are directly dependent on real interactions in ground and excited state.

## Summary

In this work, we investigate the isotopic dependence of collisional widths and shifts of the Hg clock transition $^1S_0$–$^3P_0$, when perturbed by the Rb atoms with collisional energies $E/k_B$ in the range from 1 nK to 1 K. We demonstrate the connection between the dependence of collisional line shape parameters on the reduced mass of the colliding partners, as well as the variation of the scattering length in the ground and excited state of the Hg-Rb system when elastic collisions are taken into account. The collisional width and shift of a clock transition at ultra-low temperatures can vary by orders of magnitude and exhibit resonance structure. These variations decrease at higher temperatures. However, even at a temperature around 1 K, the variations remain in the order of tens of percent. This behavior is qualitatively distinct from the isotopic effects observed at room temperature, where the expected variations are not larger than a few percent. We have indicated that the inelasticity of collisions caused by Penning ionization can change the isotopic variation of collisional broadening and shifting, even leading to elimination of resonances related to the excited state due to the universal behavior of such collisions[69].

We show that at ultra-low temperatures around 1 μK, the collisional width and shift can be accurately estimated using simple analytical expressions within the *s*-wave approximation along with the scattering lengths for colliding Hg and Rb atoms in the ground ($^1S_0$–$^2S_{1/2}$) and excited ($^3P_0$–$^2S_{1/2}$) state asymptote. On the other hand, full quantum scattering calculations of collisional width and shift, performed at a scattering energy corresponding to 1 K, agree with the classical approximation within 30% and 60%, respectively.

This paper is an extension of previous theoretical studies[78], contributing to ongoing experiments focused on collisional processes in a trapped mixture of ultra-cold Hg and Rb atoms[42]. The applicability of this work for the RbHg system is straightforward, however, the insights can also be extended to other similar systems. The overall dependence of the resonance structure on collisional energy is expected to exhibit similar behavior in systems like RbYb[71,100,101], RbSr[82,101,102], CsYb[100,101] and RbCd. Thus, the results provide valuable guidance for understanding and designing experiments involving these systems.

## Data availability

The datasets generated and analyzed during the current study are available in the open repository[103].

## Acknowledgements
We thank Mateusz Borkowski for the fruitful discussions. The research was performed at the National Laboratory FAMO (KL FAMO) in Toruń, Poland, and was supported by a subsidy from the Polish Ministry of Science and Higher Education. Some of the computations reported in this work were performed using resources of the prime cluster at the Institute of Physics, NCU.

## Author contributions
R.B. developed the code and prepared all the figures, A.L. implemented thermal averaging using Maxwell-Boltzmann distribution to the code, R.B. and A.L. performed the numerical calculations, P.S.J, P.S.Ż and R.C. developed methodology, P.S.Ż conceptualized potential energy curves, M.W. and M.Z. set experimental context of the project, R.C. conceptualized and supervised the project, R.B., A.L., M.W., P.S.Ż, M.Z., R.C., and P.S.J. analyzed the results. All authors wrote, reviewed, and approved the manuscript.

## Funding
Renu Bala, Piotr S. Żuchowski and Roman Ciuryło acknowledge Polish National Science Centre Project No. 2021/41/B/ST2/00681 support. Adam Linek acknowledges Polish National Science Centre Project No. 2023/49/N/ST2/03620 support. Marcin Witkowski acknowledges Polish National Science Centre Project No. 2021/42/E/ST2/00046 support. We acknowledge funding from the EURAMET European Partnership on Metrology 23FUN02 CoCoRICO project, co-financed from the European Union's Horizon Europe Research and Innovation Programme and by the Participating States.

## Declarations

## Competing interests
The authors declare no competing financial interests.

## Additional information
**Correspondence** and requests for materials should be addressed to R.B.

**Reprints and permissions information** is available at www.nature.com/reprints.

**Publisher's note** Springer Nature remains neutral with regard to jurisdictional claims in published maps and institutional affiliations.